\documentclass{aa}
\usepackage{txfonts}
\usepackage{graphicx}

\def\msun{\hbox{M$_\odot$}}

\begin{document}

\title{Discovery of rotation axis alignments in Milky Way globular clusters}

\author{Andr\'es E. Piatti\inst{1,2}\thanks{\email{andres.piatti@unc.edu.ar}}}

\institute{Instituto Interdisciplinario de Ciencias B\'asicas (ICB), CONICET-UNCUYO, Padre J. Contreras 1300, M5502JMA, Mendoza, Argentina;
\and Consejo Nacional de Investigaciones Cient\'{\i}ficas y T\'ecnicas (CONICET), Godoy Cruz 2290, C1425FQB,  Buenos Aires, Argentina\\
}

\date{Received / Accepted}

\abstract{
There is an increasing number of recent observational results which show that 
some globular clusters exhibit internal rotation while they travel along their orbital 
trajectories around the Milky Way center. Based on these findings, we looked for
any relationship between the inclination angles of the globular clusters' orbits with
respect to the Milky Way plane and those of their rotation. We discovered that the
relative inclination, in the sense rotation axis inclination $-$ orbit axis inclination,
is a function of the  globular cluster's orbit inclination. Rotation and orbit axes are
aligned for an inclination of $\sim$ 56$\degr$, while the rotation axis inclination is
far from the orbit's one between $\sim$ 20$\degr$ and $-$20$\degr$ when the 
latter increases from 0$\degr$ up to 90$\degr$. We further investigated the origin
of  such a linear relationship and found no correlation with the semimajor axes and 
eccentricities of the globular clusters' orbits, nor with the internal rotation strength, 
the globular clusters' sizes, actual and tidally disrupted masses, half-mass 
relaxation times, among others. The uncovered relationship will impact on the 
development of numerical simulations of the internal rotation of globular clusters, 
on our understanding about the interaction of the globular clusters with the
Milky Way gravitational field, and on the observational campaigns for increasing 
the number of studied globular clusters with detected internal rotation.}
 

\titlerunning{}

\authorrunning{Andr\' es E. Piatti }

\maketitle

\markboth{Andr\' es E. Piatti: }{Milky Way globular clusters' rotation axis}

\section{Introduction}

\citet{piatti2019} examined the inclinations ($I$) of globular clusters' orbits
around the Milky Way center and found that they do not align in
the same planar configuration of the dubbed Vast Polar Structure \citep{pk2014}; 
his findings was recently confirmed by \citet{rs2020}. Furthermore, he found out 
a linear 
relationship between $I$ and the eccentricity of prograde orbits (globular clusters 
rotating in the direction of the Milky Way's rotation), in the sense that the larger 
the eccentricity the higher the inclination, with a variation at most of $\sim$
10$\degr$ around the mean value at a fixed eccentricity. The unveiled correlation turned out to be also a function of the semimajor axis (or averaged Galactrocentric
 distance), in such a way that outermost globular clusters have orbits with the 
highest $I$ values and large eccentricities. As far as globular clusters with 
retrograde orbits are considered, there is mostly scatter over the whole $I$ range
 and eccentricities larger than $\sim$ 0.4.

While orbiting around the Milky Way center, globular clusters are also subject of
tidal forces, which cause they lose stars that can amount up to nearly half of their
initial masses, depending on the shape of their orbits, peri and apogalactocentric
distances, etc. Because of the redistribution of stars within the globular clusters, 
the internal dynamical evolution is also altered with respect to the expected one,
if the globular clusters were evolved in isolation. The result is that their internal
evolution is accelerated, so that globular clusters are seen at more advanced
internal dynamics evolutionary stages 
\citep[][and reference therein]{piattietal2019b}.

There is evidence that some globular clusters still keep some level of their
primordial rotation \citep{sollimaetal2019}, which has long been supposed to start 
with a complete alignment (coplanarity) with the axis of the orbital angular velocity
vector \citep[][and references therein]{tiongcoetal2018}. However, considering the
long-term interaction between globular clusters and the Milky Way potential
mentioned above,
we wonder whether the orbit and rotation planes are somehow linked, or whether 
both galactic and rotation motions are independent one to each other.
The topic has not received much of our attention in the available literature, 
possible because of the lack of a statistical significant sample of observational 
results on the rotation of globular clusters.

Precisely, in Section 2 we computed the rotation axis inclinations ($i$) of a sample
of globular clusters with respect to the Milky Way plane, in order to compare them
with the $I$ values. Section 3 deals with the analysis of such a comparison and
discusses the revealed interplay between $I$ and $i$. Finally, Section 4 
summarizes the main conclusions of this work.

\section{Rotation axis inclination estimates}

We made use of the rotation parameters derived by 
\citet[][their Table\,2]{sollimaetal2019} for the largest studied sample of globular 
clusters with detected rotation (see also their Table\,3). They fitted the amplitude 
($A$), the inclination of the rotation axis with respect to the LOS ($inc$),
and the position angle $\theta_0$ of the projected  rotation axis on the sky.
Thus, they provided with equations to compute the velocity components
in direction parallel and perpendicular to the projected rotation axis on the sky
and along the LOS of any point located at a small distance from the globular
cluster center and at a position angle $\theta$. Our strategy consisted in rotating 
that framework around the LOS axis by an angle $\theta_0$ in order to have the 
components of the rotation velocity along the R.A. and Dec. axes. In doing this, we
used the following expressions:

\begin{multline}
4.74r_0(pmra - pmra_0) = A(cos(\theta-\theta_0)cos(\theta)cos(inc) \\ -sin(\theta-\theta_0)sin(\theta_0)cos(inc)),
\end{multline}
\begin{multline}
4.74r_0(pmdec - pmdec_0) = A(sin(\theta-\theta_0)cos(\theta_0)cos(inc) \\ + cos(\theta-\theta_0)sin(\theta_0)cos(inc)),
\end{multline}
\begin{equation}
RV - RV_0= Asin(\theta-\theta_0)sin(inc);
\end{equation}

\noindent where $r_0$, $pmra_0$, $pmdec_0$, and $RV_0$ are the mean 
globular cluster heliocentric distance, R.A. and Dec. proper motions, and
radial velocity, respectively \citep{baumgardtetal2019}. We generated 
for each globular cluster a sample of points uniformly distributed along $\theta$ 
from 0$\degr$ up to 360$\degr$ ($\theta$ is measured from the North to the West) in steps of 1$\arcsec$ and at a distance from the globular clusters' centers equal 
to 10$^{-3}$ times their tidal radii \citep{baumgardtetal2019}. We then computed 
Galactic coordinates $(X,Y,Z)$ and space velocities $(V_X,V_Y,V_Z)$ employing 
the \texttt{astropy} \footnote{https://www.astropy.org} package 
\citep{astropy2013,astropy2018}, which simply required the input of $r_0$,
$pmra$, $pmdec$, and $RV$ given by eqs. (1)-(3). 

The components of the rotational angular momentum were calculated according 
to:

\begin{equation}
L_X  =  (Y - Y_0)\times (V_Z - V_{Z_0}) - (Z - Z_0)\times (V_Y - V_{Y_0}),
\end{equation}
\begin{equation}
L_Y = (Z - Z_0)\times (V_X - V_{X_0}) - (X - X_0)\times (V_Z - V_{Z_0}),
\end{equation}
\begin{equation}
L_Z = (X - X_0)\times (V_Y - V_{Y_0}) - (Y - Y_0)\times (V_X - V_{X_0}),
\end{equation}

\noindent and the inclination of the rotation axis:

\begin{equation}
i = acos\left(\frac{L_Z}{\sqrt{L_X^2 + L_Y^2 + L_Z^2}}\right);
\end{equation}

\noindent where the subscript 0 (zero) in eqs. (4)-(6) refers to the mean orbital 
position and motion of the globular cluster. We adopted as uncertainties of $i$ the 
standard deviation from all the generated individual $i$ values. Since $I$ 
\citep[see][]{piatti2019} and $i$ values range from 0$\degr$ for fully prograde 
in-plane orbits (or coplanar prograde rotation) to 90$\degr$
for polar orbits (or rotation plane perpendicular to the Galactic plane), to 
180$\degr$ for in-plane retrograde orbits or rotation planes, we defined:
$I^*$ = 180$\degr$$-$$I$ and $i^*$ = 180$\degr$$-$$i$, which resulted useful
in the subsequent analysis.

\section{Analysis and discussion}

\begin{figure}
\includegraphics[width=\columnwidth]{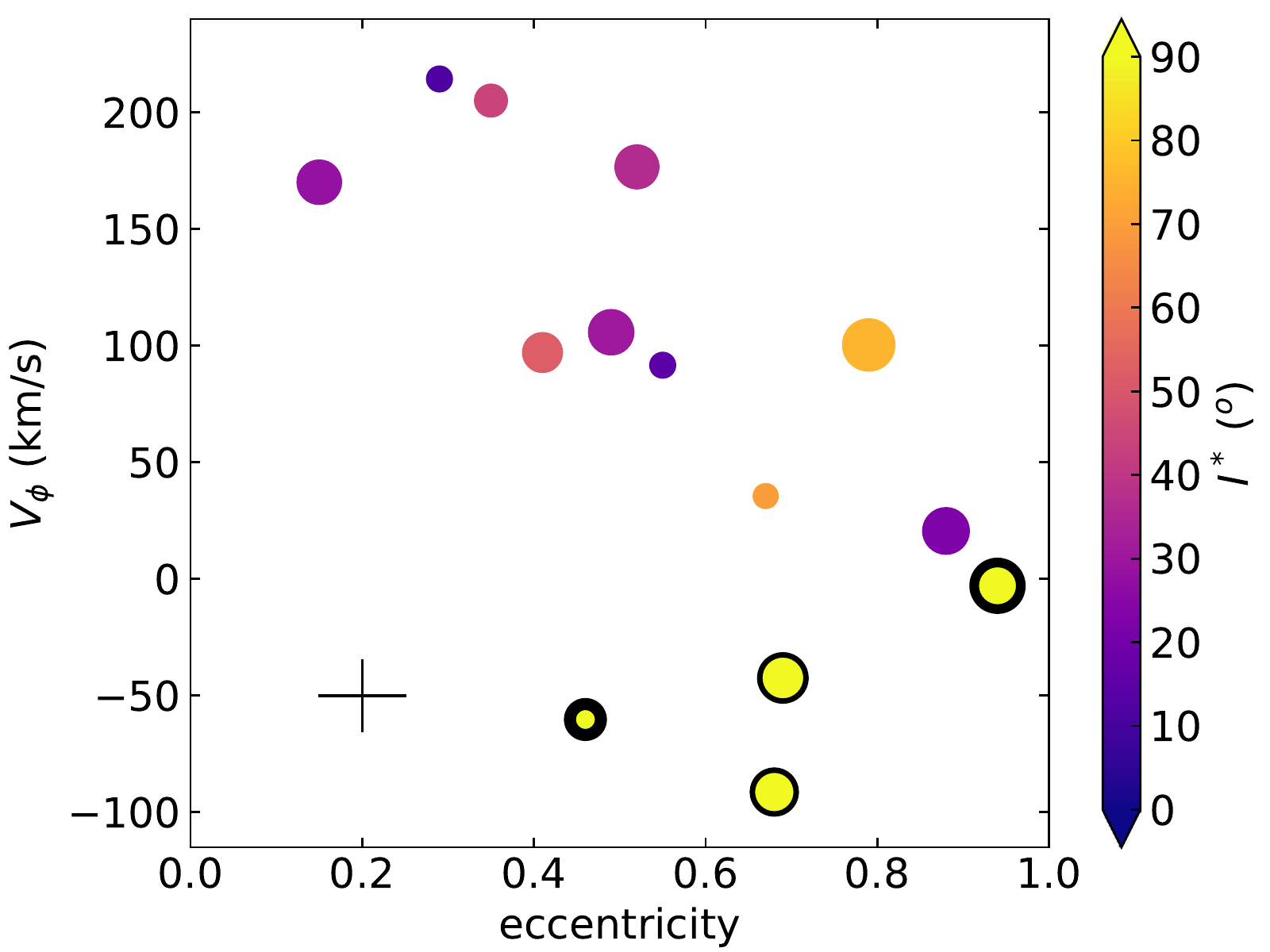}
\caption{Relationship between globular clusters' orbital parameters. Typical 
error bars are included. black-edged circles represent globular clusters with
retrograde motions. The size of the symbols is proportional to log($a$); 
the smallest and largest ones correspond to $a$= 1.6 and 13.6 kpc, 
respectively.}
\label{fig:fig1}
\end{figure}

The orbital properties of the studied globular clusters are pictured in 
Figure~\ref{fig:fig1}, where the relationship between the Galactic velocity
component $V_\phi$ (spherical coordinates), the eccentricity, and $I^*$
derived by \citet{piatti2019} is shown. As can be seen, globular clusters with 
prograde ($V_\phi$ $>$ 0) and retrograde ($V_\phi$ $<$ 0) orbits are clearly
distinguished. The latter have relatively large eccentricities and nearly polar 
trajectories ($I^*$ $>$ 80$\degr$). These globular clusters have  long been 
thought to have an accreted origin \citep{fb2010}. They have been highlighted 
in Figure~\ref{fig:fig1} with black-edged circles. Among globular clusters with 
prograde orbital motions, a trend between $V_\phi$ and the eccentricity 
arises, so that the higher the rotational velocity, the more circular their orbits 
(smaller eccentricities) and smaller $I^*$ values. Those with more circular 
orbits have tightly copied the rotation of the Milky Way disk.

We also used the semimajor axes of the globular clusters' orbits ($a$) defined 
as the average between the peri and apogalactocentric distances computed by  \citet{baumgardtetal2019}. They better represent  the distance of the
globular clusters' birthplaces to the Milky Way center or the average distance
where globular clusters were deposited after accretion of their host dwarf 
galaxy onto the Milky Way. The globular clusters in our sample span $a$ 
values from 1.6 up to 13.6 kpc, i.e., they populate the Milky Way bulge 
\citep[Galactocentric distance $<$ 3 kpc,][]{barrosetal2016} and disk. They are
represented in Figure~\ref{fig:fig1} with circles whose sizes are proportional to 
log($a$ /kpc).  Both globular clusters with prograde and retrograde orbits cover the entire range of $a$ values. From this point of view, the studied globular 
clusters can be considered as representative of the whole globular cluster population in the Galactic volume considered. Therefore, if some relationship 
existed between  $I*$ and $i*$, this should be discovered from them.

We  firstly calculated the mean and dispersion of $i^*$ by employing a 
maximum likelihood approach. The relevance lies in accounting for the 
individual $i^*$ measurement errors, which could artificially inflate the 
dispersion if ignored. We thus optimized the probability $\mathcal{L}$ that a
given ensemble of $i^*$ values with errors $\sigma$($i^*$) are drawn from a population with mean rotation axis inclination $<$$i^*$$>$ and intrinsic 
dispersion W  \citep[e.g.,][]{pm1993,Walker2006}, 
as follows:

\begin{equation}
\mathcal{L}\,=\,\prod_{k=1}^N\, \left(\,2\pi\left[\sigma_k^2 + W^2 \, \right]\,\right)^{-\frac{1}{2}} 
\times\,\exp \left(-\frac{\left(i^*{_k} \,- <i^*>\right)^2}{\sigma_k^2 + W^2} \right), 
\end{equation}

\noindent where the errors on the mean and dispersion were computed from 
the respective covariance matrices.  We would like to note that this approach 
assumes that the error distribution is Gaussian, which is adopted here because
of the limited number of globular clusters \citep[cf.][]{franketal2015}. The resulting 
mean $i^*$ and dispersion turned out to be (61.8 $\pm$ 4.8)$\degr$ and (10.5 
$\pm$ 0.7)$\degr$, respectively. This outcome reveals that the rotation axis 
inclinations are not randomly distributed -- they do not span the whole range of 
values ([0$ \degr$, 90$\degr$]) --, and are not all aligned to the orbit axes, but 
span a moderately small intrinsic range. Since the mean $i^*$ value and 
dispersion come from globular clusters that rotate in the same direction of their 
orbital motions (prograde or retrograde orbits), or in the opposite direction to their 
orbital motions, the above result could suggest that there should be some 
condition that has led the rotation axis inclinations to be  more or less polarized, 
irrespective of the direction of the orbital motion and that of the rotation. The small 
globular cluster sample analyzed could jeopardize such a speculation, if a larger
sample were used instead. However, this would not be the case, as judged by the 
subsequent results.

\begin{figure}
\includegraphics[width=\columnwidth]{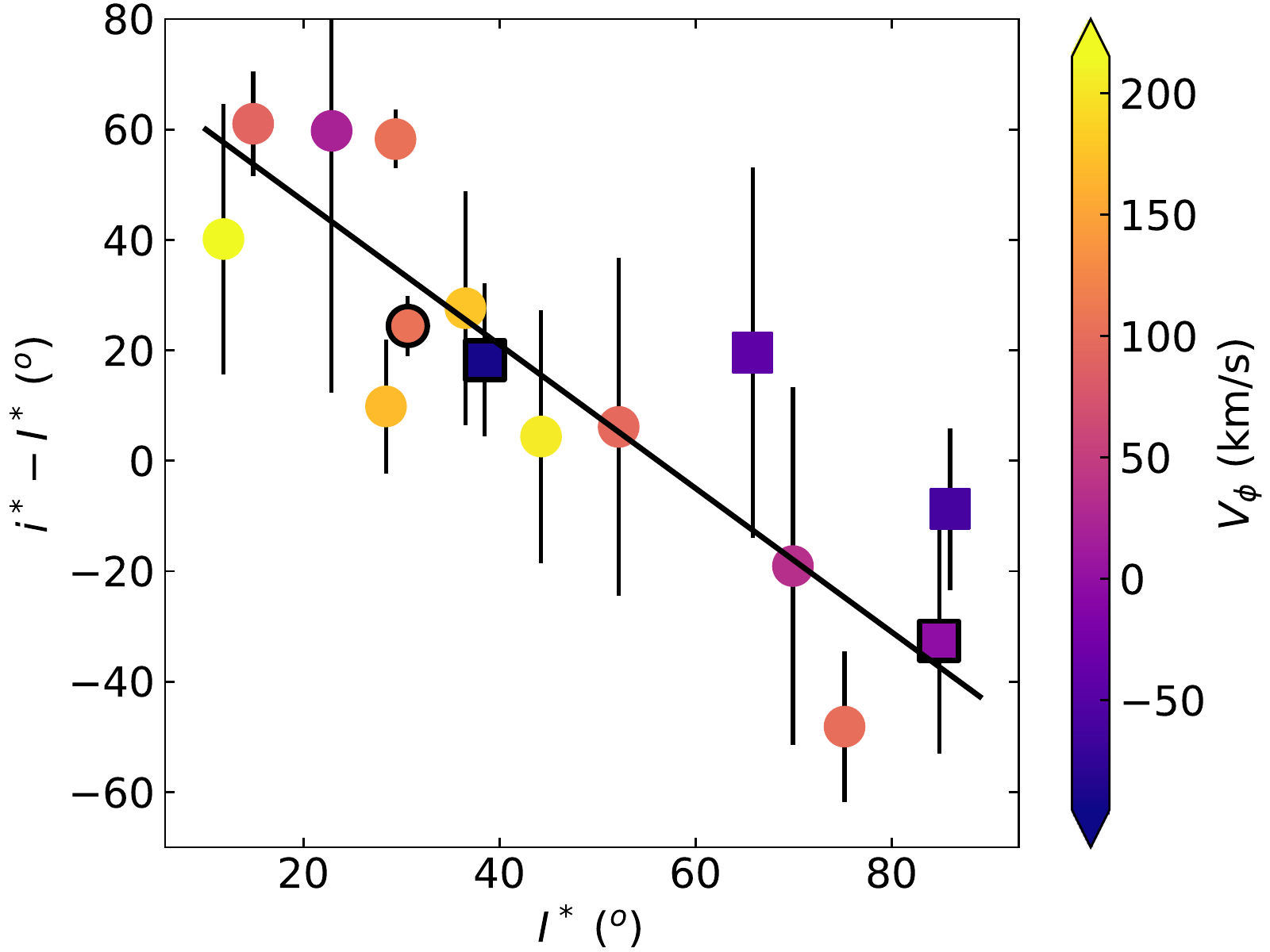}
\caption{Relative inclination versus orbital plane inclination for the studied
globular clusters. Circles and squares refer to globular clusters moving around
the Milky Way center or rotating in prograde and retrograde directions, respectively. Black-edged symbols represent globular clusters with opposite direction of orbital and rotational motions. Individual error bars are
indicated; those for $I^*$ are as large as the symbols' sizes. The black line
represent the fitted relationship (see text).}
\label{fig:fig2}
\end{figure}

Figure~\ref{fig:fig2} depicts the difference between the rotation and orbit axes'
inclinations as a function of $I^*$. We have distinguished those globular clusters 
with prograde or retrograde orbital motions, and those with prograde or retrograde
rotation. There is one globular cluster (NGC\,7078) with a prograde orbit and two 
globular clusters (NGC\,5139, 7089) with retrograde orbits that rotate in the opposite direction to their orbital motions. They have highlighted with black-edged
 symbols. The figure uncovers a relationship between $i^*$ and $I^*$ that, as far 
 as we are aware, has not been hypothesized nor found in numerical simulations 
 of the evolution of rotating star clusters, nor discovered observationally. As can be
  seen, the relative inclinations follow a linear relationship with $I^*$ that resulted
 to be:

\begin{equation}
i^* - I^* = (-1.30 \pm 0.16)I^* + 73.00 \pm 6.53,
\end{equation}

\noindent with a standard deviation of $\pm$18.2$\degr$, when we fitted the
points by least square. We obtained a correlation of 0.80 and an $F$-test 
coefficient of 0.90, which undoubtedly rejects the possibility that the relative 
inclinations for the small sample of 15 globular clusters analyzed were 
dominated by a point dispersion.

This finding reveals that if the rotation axis was aligned to the orbit axis at the 
globular cluster's birth, such a  configuration changed in the long-term evolution. 
Otherwise, if the rotation axis is thought of not to change along the globular 
cluster's lifetime, then the traditional scenario of coplanarity between the orbit and the rotation's planes as initial condition for the globular cluster evolution would not 
longer be supported \citep{tiongcoetal2018}. According to eq. (9), present-day
coplanarity is observed in globular clusters with an orbit inclination  $I^*_{0}$ = 
$i^*_{0}$  $\sim$ 56$\degr$. Globular clusters with orbit inclinations smaller and 
larger than $I^*_{0}$ rotate around an axis inclined with respect to the Milky Way plane that varies up to $\sim$ 20$\degr$ in excess or defect around $i^*_{0}$, respectively. Eq. (9) also allows us to predict the rotation axis inclination of a
globular cluster, provided that the inclination of its orbit is known.

\begin{figure*}
\includegraphics[width=\textwidth]{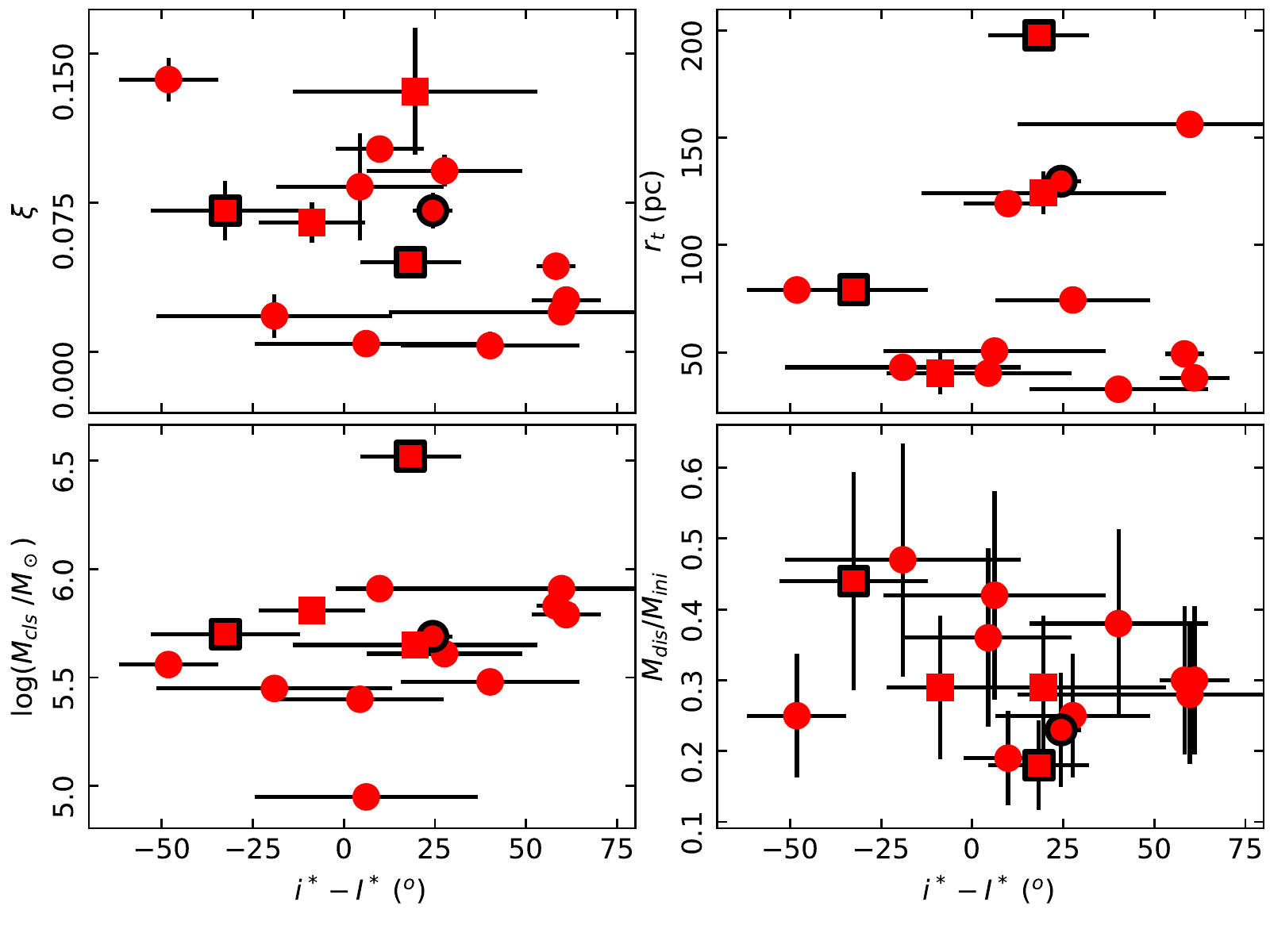}
\caption{ Relative inclinations as a function of different
astrophysical properties.}
\label{fig:fig3}
\end{figure*}

We tried to find out some trail of the physical origin of eq. (9) by  examining
the relationships of $i^* - I^*$ with different 
globular clusters' parameters taken from \citet{piatti2019} and 
\citet{piattietal2019b}, if another source is not mentioned. We considered $a$  and
the orbit eccentricity to link eq. (9) to orbital properties; the present mass
(log($M_{cls}$ /$\msun$)), the ratio between the mass lost by tidal disruption to 
the total mass ($M_{dis}/M_{ini}$) and the ratio of the age to the half-mass 
relaxation time to see whether the internal dynamical evolution has a role; the tidal
radius ($r_t)$ and the rotation strength \citep[$\xi,$][]{sollimaetal2019} to connect 
eq. (9) to the internal rotation itself, and the status of a globular cluster according
to whether it has tidal tails, extra-tidal features different to tidal tails, or simply a 
\citet{king62}'s profile \citep{pcb2020}. Figure~\ref{fig:fig3} shows some of these 
plots. As can be seen, none of them would seem to make clear that eq. (9) 
correlates with any of these globular cluster's astrophysical properties. 
For this reason, we speculate with the possibility of suggesting looking for the
origin of the interplay between $i^*$ and $I^*$ in some large-scale effect
caused by Milky Way's characteristics, such as its gravitational potential, 
overall magnetic field, among other.

\section{Concluding remarks}

Motivated by some recent observational results which have presented reliable 
globular clusters' orbital and rotation parameters, we sought for any connection
between the inclination angle of the orbital plane and that of the rotation with
respect to the Milky Way plane. We analyzed 15 globular clusters, which at 
present is the largest sample of globular clusters with comprehensive rotation
studies.

Our analysis relied on the recently published rotation parameters referred to the
plane of the sky and the LOS as the natural framework. From them, we 
transformed the rotation velocity components to the Galactic coordinate system 
and computed both the inclination angle of the globular clusters' orbits and those
of their rotation. We found that the present-day globular clusters' rotation axis 
inclinations are not aligned with those of the globular clusters' orbits nor with the 
Galactic poles one. They resulted to be inclined with respect to the corresponding 
orbit axis inclination, varying from $\sim$ 60$\degr$ to $\sim$ -40$\degr$,
when the latter increases from 0$\degr$ up to 90$\degr$. Further investigations
are needed in order to find out the cause of such a behavior.

The linear relationship found between the relative inclinations and the orbits' 
inclinations provides new observational evidence that will impact on the future
numerical simulations of the internal globular clusters' dynamics, on our 
understanding of the interaction of globular clusters with the Milky Way, on
the increase of the number of studied globular clusters with derived rotations.
From the 156 globular clusters cataloged in \citet[][2010 Edition]{harris1996}
62 have been studied by \citet{sollimaetal2019}, and from them 15 resulted
with non-negligible internal rotation. This means that there should be nearly
22 more globular clusters in \citet{harris1996} with detectable internal rotation.

\begin{acknowledgements}
I thank Pavel Kroupa and Antonio Sollima for preliminary discussions
about the rotation of globular clusters.
I also thank the referee for the thorough reading of the manuscript and
timely suggestions to improve it. 
\end{acknowledgements}



\end{document}